\documentclass[twocolumn]{jpsj2} 
\title{Correlation between Superconducting Transition Temperature $T_c$ and Increase of Nuclear Spin-Lattice Relaxation Rate Devided by Temperature $1/T_1T$ at $T_c$ in the Hydrate Cobaltate Na$_{x}$CoO$_{2}\cdot y$H$_{2}$O}
\author{
Y. \textsc{Ihara,}\thanks{E-mail address: ihara@scphys.kyoto-u.ac.jp}
K. \textsc{Ishida,}\thanks{E-mail address: kishida@scphys.kyoto-u.ac.jp}
C. \textsc{Michioka,}$^{1}$
M. \textsc{Kato,}$^{1}$
K. \textsc{Yoshimura,}$^{1}$
K. \textsc{Takada,}$^{2}$
T. \textsc{Sasaki,}$^{2}$
H. \textsc{Sakurai,}$^{3}$
and
E. \textsc{Takayama-Muromachi,}$^{3}$}

\inst{Department of Physics, Graduate School of Science, Kyoto University, Kyoto 606-8502, Japan \\
$^{1}$Department of Chemistry, Graduate School of Science, Kyoto University, Kyoto 606-8502, Japan \\
$^{2}$Advanced Materials Laboratory, National Institute for Materials Science, 1-1 Namiki, Tsukuba Ibaraki 305-0044, Japan\\
$^{3}$Superconducting Materials Center, National Institute for Materials Science, 1-1 Namiki, Tsukuba Ibaraki 305-0044, Japan}

\abst{We have performed Co-nuclear quadrupole resonance (NQR) studies on Na$_{x}$CoO$_{2}\cdot y$H$_{2}$O compounds with different Na ($x$) and hydrate ($y$) contents. Two samples with different Na contents but nearly the same $T_c$ values ($x$ = 0.348, $T_c$ = 4.7 K ; $x$ = 0.339, $T_c$ = 4.6 K) were investigated. The spin-lattice relaxation rate $1/T_1$ in the superconducting (SC) and normal states is almost the same for the two samples except just above $T_c$. NQR measurements were also performed on different-hydrate-content samples with different $T_c$ values, which were prepared from the same Na-content ($x$ = 0.348) sample. From measurements of $1/T_1$ using the different-hydrate-content samples, it was found that a low-$T_c$ sample with $T_c = 3.9$ K has a larger residual density of states (DOS) in the SC state and a smaller increase of $1/T_1T$ just above $T_c$ than a high-$T_c$ sample with $T_c$ = 4.7 K. The former behavior is consistent with that observed in unconventional superconductors, and the latter suggests the relationship between $T_c$ and the increase in DOS just above $T_c$. This increase, which is seemingly associated with the two-dimensionality of the CoO$_2$ plane, is considered to be one of the most important factors for the occurrence of superconductivity.  }

\kword{unconventional superconductivity, hydrate sodium cobaltate, NQR, spin fluctuations}

\begin{document}
\maketitle
Since the discovery of the layered sodium cobaltate superconductor Na$_{x}$CoO$_{2}\cdot y$H$_{2}$O,\cite{Takada} much attention has been paid to the magnetic properties of the CoO$_{2}$ plane with a triangular lattice.
The CoO$_2$ forms a two-dimensional (2D) hexagonal layered structure and, in the nominal notation, the $1-x$ fraction of Co is in the low spin ($S$ = 1/2) Co$^{4+}$ state while the $x$ fraction of Co is in the $S$ = 0 Co$^{3+}$ state. 
Superconductivity was discovered in Na$_{0.35}$CoO$_2$ $\cdot 1.3$H$_2$O, in which the sodium concentration is reduced to 0.35 and an intercalation of H$_2$O occurs.

The above cobaltate superconductivity is quite interesting compared with cuprate superconductivity, since both systems commonly have partially filled 3$d$ orbitals. In cuprates, Cu$^{2+}$ has an $e_g$($d_{x^2-y^2}$) band near the Fermi level with a strong hybridization with O 2$p$ orbitals, whereas in cobaltates, $3d^5$ in Co$^{4+}$ occupies three lower $t_{2g}$ bands with less hybridization to O 2$p$ orbitals due to the weaker overlap of triangular bonding.   
At present, the magnetic properties of cobaltate superconductors are controversial. It is believed that a cobaltate is also an electron-doped Mott insulator the same as an electron-doped cuprate, since 0.35 Na doping in CoO$_2$ is considered to provide 0.35 electrons/Co in a frustrated half-filled triangular lattice.
On the other hand, Na$_{0.35}$CoO$_2$ can also be regarded as 0.65 hole doping/Co to the band-insulating state in NaCoO$_2$ with the low-spin state of Co$^{3+}$.
The former situation is favorable for the resonating valence bond (RVB) state, which prefers a $d$-wave order parameter\cite{Baskaran,Ogata}, and the latter situation in which the doping level is far from the half filling on the 2D triangular lattice can give rise to $f$-wave triplet superconductivity.\cite{Kuroki,Ikeda}
The proper understanding of the magnetic properties of the CoO$_2$ plane is quite important in elucidating the mechanism of this superconductivity.
In addition, it is reported that the superconducting (SC) phase diagram of Na$_x$CoO$_2 \cdot y$H$_2$O has a dome shape with respect to $T_c$ against $x$, which is similar to that of cuprate superconductors.\cite{Schaak} Investigation into the Na-doping dependence of $T_c$ and the magnetic properties of the CoO$_2$ plane is also important for understanding this system, however there are few studies on these points.

In our previous paper, from the measurement of spin-lattice relaxation rate ($1/T_1$) in Na$_{0.348}$CoO$_2 \cdot$ 1.3H$_2$O with $T_c = 4.7$ K by the nuclear quadrupole resonance (NQR) technique, we reported that the coherence peak is absent just below $T_c$, accompanied with the $T_1T$ = const.(Korringa) behavior far below $T_c$. This strongly suggests that this superconductivity is classified as unconventional superconductivity from the $1/T_1$ behavior in the SC state.\cite{Ishida}
We also reported that $1/T_1T$ is enhanced below 100 K with decreasing temperature and showed that this enhancement is related to the development of $q = 0$ fluctuations from comparison between $1/T_1T$ and $\chi_{\rm bulk}$. 
Therefore, we suggest that nearly ferromagnetic fluctuations might be associated with the unconventional superconductivity.\cite{Ishida}

To date, NQR results have been reported by several groups, however the results on $1/T_1$ behavior are not consistent.\cite{Ishida,Kobayashi,Fujimoto}
The difficulty in carrying out experiment on the hydrate-cobaltate superconductor is that water in the samples is easily removed in atmosphere and that the SC characteristics degrade rapidly. It was also reported that the SC characteristics are highly influenced by Na content.\cite{Schaak}
To settle the discrepancy in the NQR results reported thus far, systematic NQR studies using well-characterized samples are necessary.

In this paper, we report the NQR results of several samples with different Na and hydrate contents. The NQR results of two samples with different Na contents but nearly the same $T_c$ value are very similar except for the $1/T_1T$ behavior just above $T_c$. On the other hand, the NQR results of the different-hydrate-content samples prepared from same-Na-content samples suggest that $T_c$ and the electronic state in the normal state are dependent on hydrate content, which is seemingly related to the $c$-axis lattice constant. The dependence of $T_c$ on the $c$-axis lattice constant is consistent with a recent report by Milne {\it et al}.\cite{Milne} A high-$T_c$ sample with a high hydrate content shows a large increase in $1/T_1T$ in the normal state, together with a gradual increase in susceptibility below 100 K.
The present results show that $T_c$ depends on normal-state electronic state, which is dependent on hydrate content and/or $c$-axis lattice constant. 
      
\begin{figure}[tb]
\begin{center}
\includegraphics[width=6.5cm]{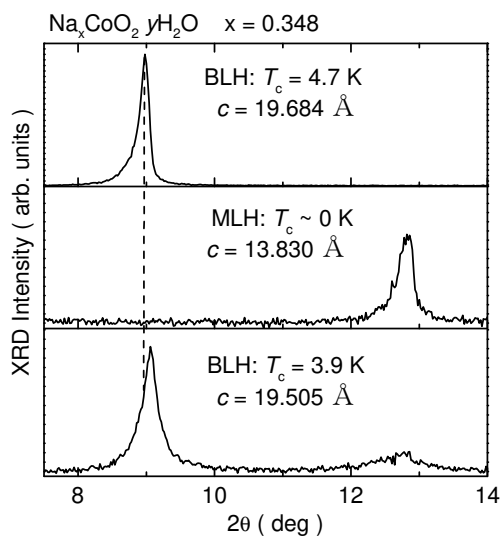}
\end{center}
\caption{X-ray diffraction of three samples with different hydrate contents for $7.5^{\circ} < 2\theta < 14^{\circ}$. The peaks shown here correspond to the (002) diffraction.}
\label{f1}
\end{figure}
In the present experiment, we used powder samples, the preparation of which has been reported previously.\cite{Takada} A SC transition was confirmed by dc-susceptibility measurement. The Na content was analyzed using inductively coupled plasma atomic-emission spectrometry (ICP-AES). Two samples with different Na contents were investigated. One was a sample with Na content $x$=0.348, which shows superconductivity at 4.7 K; the other was a sample with $x$ =0.339 and $T_c= 4.6$ K. The NQR results of the former sample were reported in our previous paper.\cite{Ishida}
The preparation of different-hydrate-content samples from the same $x$ = 0.348 sample was as follows. The fully hydrated sample with $x$ = 0.348 was stored in vacuum at room temperature for three days, and it turned into a mono-layered hydrate (MLH) structure. This MLH sample did not show superconductivity down to 1.5 K. After this MLH sample was left in a humid atmosphere for one week, the superconductivity recovers with $T_c$ = 3.9 K.\cite{Takada} We call this sample as the ``BLH (bilayered hydrate) low-$T_c$ sample'' in this paper to distinguish it from the original ``BLH high-$T_c$ sample''. In the above procedure, Na content is considered to be unchanged, thus the difference in $T_c$ is ascribed to that in hydrate content $y$ in Na$_{0.348}$CoO$_2$$\cdot$ $y$H$_2$O. 
From the X-ray diffraction (XRD) result shown in Fig.~1, the MLH sample does not include the BLH structure at all, whereas the BLH low-$T_c$ sample has a small fraction of the MLH structure. The quality of the BLH low-$T_c$ sample is not as good as the original BLH high-$T_c$ sample due to the imperfect hydration process.
It is confirmed here that the SC phenomenon is reversible for the hydration process.

\begin{figure}[tb]
\begin{center}
\includegraphics[width=6.5cm]{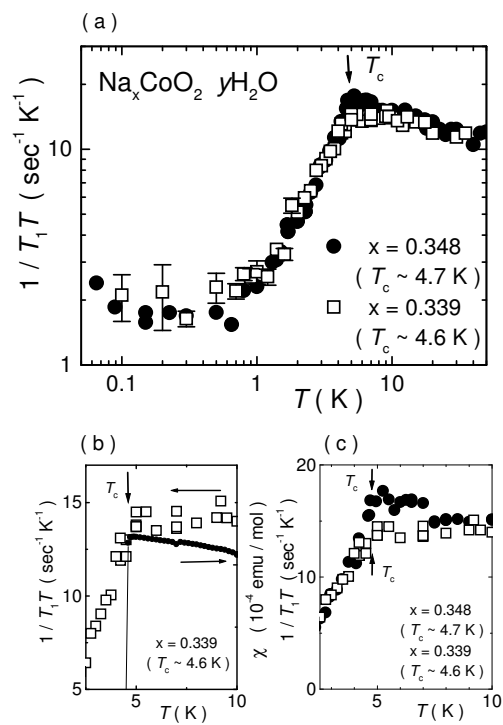}
\end{center}
\caption{(a) Temperature dependence of $1/T_1T$ of different-Na-content samples measured at NQR peak of 12.3 MHz. (b) Comparison between $1/T_1T$ and $\chi$. $1/T_1T$ starts to decrease just below the SC transition temperature determined by $\chi$. (c) Temperature dependence of $1/T_1T$ of two samples at approximately $T_c$. $T_c$ is denoted by an arrow in each figure.}
\label{f2}
\end{figure}
First we discuss the NQR results of the two different-Na-content samples ($x$ = 0.348, $T_c$ = 4.7 K: $x$ = 0.339, $T_c$ = 4.6 K). NQR spectra of the two samples arising from the $\pm 7/2 \leftrightarrow \pm 5/2$ transition ($^{59}$Co: $I$ = 7/2) show a maximum at 12.3 MHz with a full width at half maximum (FWHM) of $\sim 0.4$ MHz. A difference in NQR spectra is not observed between the two samples.
Figure 2(a) shows the temperature dependence of $1/T_1T$ in the $x$ = 0.348 and $x$ =0.339 samples, measured at the maximum peak in the NQR spectra. $1/T_1T$ in both samples shows a sharp decrease without a coherence peak just below $T_c$, accompanied by the Korringa relation far below $T_c$ due to the residual density of states (DOS) at the Fermi level. Detailed comparison between $1/T_1T$ and $^{\rm dc}\chi$ at approximately $T_c$ in the $x$ = 0.339 sample is shown in Fig.~2(b). It is also confirmed in the $x$ = 0.339 sample that $1/T_1T$ starts to decrease at $T_c$ determined by $^{\rm dc}\chi$.
Figure 2(a) shows that the residual DOS determined by the Korringa relation is almost the same for the two samples. The close magnitudes of the residual DOS are consistent with the very close $T_c$ values of the two samples, since it is well established that the residual DOS has a relationship with $T_c$ in unconventional superconductivity, as will be discussed later. A small difference in normal-state $1/T_1T$ between the two samples was found below 10 K as shown in Fig.~2(c). $1/T_1T$ in the $x$ = 0.348 sample increases with decreasing $T$, whereas it is constant below 10 K in the $x$ = 0.339 sample. The possible origin of this difference in $1/T_1T$ in the normal state will be discussed later.

\begin{figure}[tb]
\begin{center}
\includegraphics[width=6.5cm]{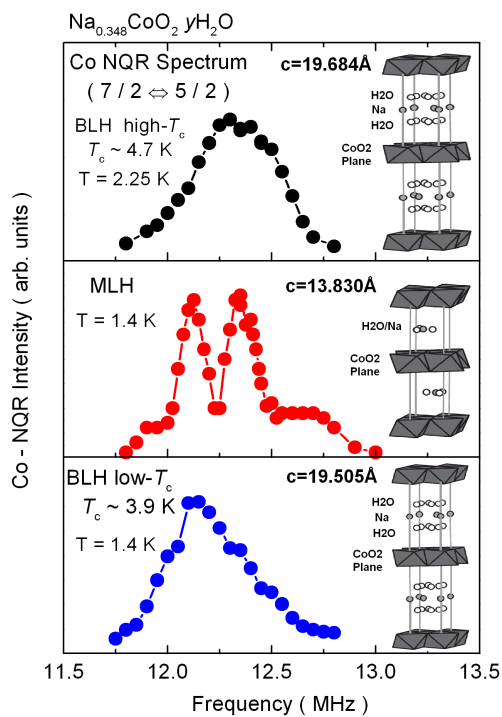}
\end{center}
\caption{$^{59}$Co-NQR spectra corresponding to $\pm5/2 \leftrightarrow \pm7/2$ transition in BLH high-$T_c$, MLH, and BLH low-$T_c$ samples. The spectra were obtained by a frequency-swept method. The crystal structure of each compound is also shown for comparison.  }
\label{f1}
\end{figure}
Next, we report the magnetic properties of the different hydrate-content samples.
Figure 3 shows the Co-NMR spectra of the different hydrate-content samples, which arise from the $\pm 5/2 \leftrightarrow \pm7/2$ (3$\nu_Q$) transition. The crystal structure of each compound is also shown in Fig.~3. The BLH high-$T_c$ sample shows a single peak at 12.3 MHz as discussed above, and the MLH shows two sharp peaks at 12.1 and 12.3 MHz together with other small satellite peaks at lower and higher frequencies. The BLH low-$T_c$ sample shows a peak at 12.1 MHz with FWHM $\sim$ 0.4 MHz. In the MLH sample, there exist several Co sites with different Na configurations around a Co ion since Na and H$_2$O are in the same layer. The complex spectrum with the multipeak structure reveals the existence of such Co sites in different environments. The identification of each peak is now in progress.
On the other hand, both of the SC BLH samples show a single peak as shown in Fig.~3. This shows that the random effect on the Co site, which originates from the partial occupation of the Na layer, is screened by the two adjacent H$_2$O layers.  
It seems that the unique electric field gradient at the Co sites is related to the occurrence of superconductivity. The frequency difference between the two BLH samples is considered to originate from the difference in the distance between the Co and hydrate ions, since the XRD measurements reveal that the $c$-axis lattice parameter in the BLH low-$T_c$ sample is shorter than that in the BLH high-$T_c$ sample as shown in Fig.~1.\@ 

$T_1$ was measured at the maximum peak in the BLH samples, and at 12.1 and 12.3 MHz in the MLH sample. At these peaks, the recovery of nuclear magnetization after saturation pulses can be satisfactorily fitted to a theoretical curve\cite{MacLaughlin} in the whole temperature range except at very low temperatures.

\begin{figure}[tb]
\begin{center}
\includegraphics[width=7cm]{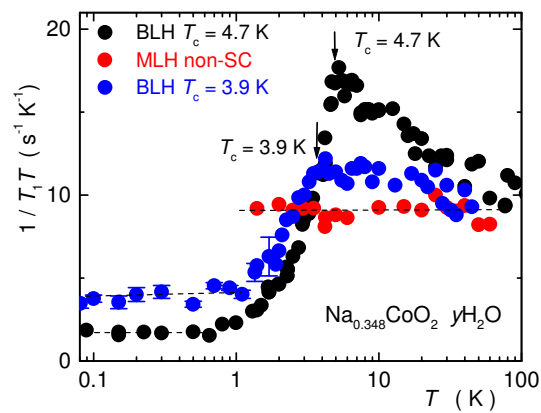}
\end{center}
\caption{Temperature dependence of $1/T_1T$ in different-hydrate-content samples. Temperature is plotted on the logarithmic scale}
\label{f1}
\end{figure}
\begin{figure}[t]
\begin{center}
\includegraphics[width=6.5cm]{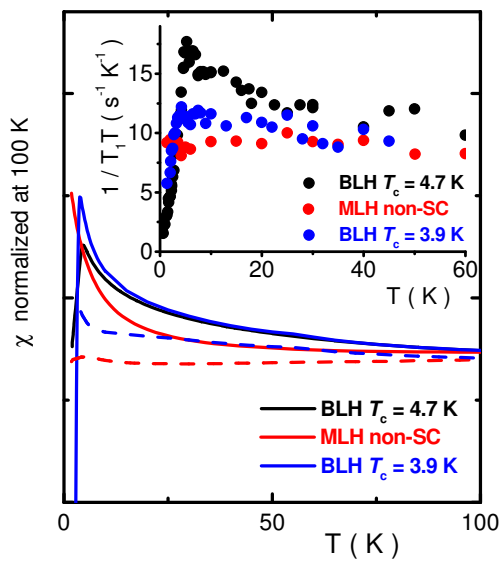}
\end{center}
\caption{Temperature dependence of bulk susceptibility $\chi$ normalized to that at 100 K in different-hydrate-content samples. The inset shows $1/T_1T$ in these samples plotted against $T$ on a linear scale. The red (blue) broken curve shows the $T$ dependence of $\chi$ in the MLH (BLH low-$T_c$) sample after the Curie tail of the MLH was subtracted.  }
\label{f1}
\end{figure}

Figure 4 shows the temperature dependence of $1/T_1T$ measured in these samples.
In the MLH sample, $1/T_1T$ measured at 12.3 MHz is shown in Fig.~4.\cite{MLH}
$1/T_1T$ in the MLH sample shows the Korringa behavior below 100 K down to 1.9 K, whereas $1/T_1T$ in the two SC BLH samples shows a sharp decrease below $T_c$. The decreasing rate of $1/T_1T$ just below $T_c$ is faster in the high-$T_c$ sample than in the low-$T_c$ sample. This suggests that the sharp decrease is the intrinsic behavior of the system. If the residual DOS is estimated from the Korringa behavior in the SC state, the residual DOS of the BLH low-$T_c$ sample is larger than that of the BLH high-$T_c$ sample.
This relation between the residual DOS and $T_c$ has been reported in various unconventional superconductors such as cuprate\cite{Hotta,IshidaZn} and Sr$_2$RuO$_4$\cite{IshidaSRO} superconductors.
In the normal state, $1/T_1T$ in the BLH high-$T_c$ sample continues to increase below 100 K down to $T_c$, whereas that in the BLH low-$T_c$ sample is constant below 20 K.
We consider that the increase in $1/T_1T$ below 100 K depends on $T_c$, and that this increase is important for the occurrence of superconductivity.

Figure 5 shows the temperature dependence of the bulk susceptibility $^{\rm dc}\chi$ measured at 10 kOe.
Due to the lack of the exact hydrate content $y$ and nearly temperature-independent behavior above 100 K, $\chi$ is normalized to that at 100 K.
The $\chi$ in the MLH sample is nearly constant down to 25 K and increases sharply below 25 K, whereas the $\chi$ in the SC samples increases gradually below 100 K. It seems that there are two origins for the increase in $\chi$. The gradual increase below 100 K, which is observed in the two SC samples, is related to the Curie-Weiss behavior of $1/T_1T$.
On the other hand, the sharp increase below 25 K, which is observed in the MLH and BLH low-$T_c$ samples, is ascribed to some phase other than Na$_{x}$CoO$_2 \cdot y$H$_2$O. If we take into account the fact that this sharp increase is not observed at all in the $1/T_1T$ behavior, this sharp increase is due to an impurity phase with a small volume fraction, {\it e.g.}, Na$_{x}$CoO$_2$ without water, which arose in the dehydration process. This is because Co NQR signal in the Na$_x$CoO$_2$ without water would be found at a frequency other than $\sim$ 12 MHz \cite{Ray}.
If the Curie tail in the MLH sample is subtracted from the observed $\chi$ in the MLH and BLH low-$T_c$ samples, the $T$ dependence of the resultant $\chi$ is shown by the broken curves in Fig. 5.\@ The BLH low-$T_c$ sample shows the gradual increase below 100 K, which is weaker than that observed in the BLH high-$T_c$ sample as $1/T_1T$ does. 
We suggest that the variation in magnetic properties revealed by $1/T_1T$ and $\chi$ is closely related to the $T_c$ difference between the two BLH samples, which is seemingly induced by the difference in the $c$-axis lattice parameter of these samples, since the relationship between $T_c$ and the $c$-axis lattice parameter was observed.\cite{Milne}  

We suggested in our previous study\cite{Ishida} that nearly ferromagnetic fluctuations are dominant in the normal state, which are related to the increase in DOS at the Fermi level. Although we show here the relationship between $1/T_1T$ and $\chi$, the identification of ferromagnetic fluctuations is not sufficient since we cannot rule out the possibility of incommensurate antiferromagnetic fluctuations. However, it is reasonable to conclude that the increases of $1/T_1T$ and $\chi$ are ascribed to the increase in DOS at the Fermi level. We suggest that the increase in DOS, which could be associated with the longer distance between the CoO$_2$ layers, plays a vital role in the occurrence of superconductivity.  

Finally, we briefly comment on the difference in the $1/T_1T$ results reported by different groups.\cite{Ishida,Kobayashi,Fujimoto}
The difference in $1/T_1T$ is seen mainly in two temperature regions, {\it i.e.}, the temperature regions far below $T_c$ and that just above $T_c$. $1/T_1T$ in the former region is dominated by the residual DOS in the SC state, and $1/T_1T$ in the latter region is related to the increase in the DOS discussed above. These differences in $1/T_1T$ are considered to be mainly due to the different hydrate contents in these samples. We consider that the small difference in $1/T_1T$ below 10 K observed between the two different Na-content samples is also due to the slight difference in their hydrate content, since the tendency of the sharp increase in $\chi$ seen in hydrate-deficient samples below 25 K is observed in the $x$ = 0.339 sample.

In conclusion, from the Co-NQR studies on Na$_{x}$CoO$_{2}\cdot y$H$_{2}$O compounds with different Na ($x$) and hydrate ($y$) contents, it seems that Na content is not so sensitive to $T_c$ and magnetic fluctuations.
On the other hand, we found that the BLH high-$T_c$ sample has a larger enhancement of $1/T_1T$ just above $T_c$ than the BLH low-$T_c$ sample. If we consider the longer $c$-axis lattice parameter in the BLH high-$T_c$ sample, the distance between the CoO$_2$ layers is one of the most important parameters for the occurrence of superconductivity.
It is considered that the longer distance between the CoO$_2$ layers makes the system more 2-D-like and increase DOS at the Fermi level. We suggest that this increase in DOS is closely associated with the occurrence of unconventional superconductivity. 

We thank Y.~Maeno, H.~Yaguchi and S.~Nakatsuji for experimental support and valuable discussions. We also thank H.~Ikeda, K.~Yamada, Y.~Yanase and M.~Ogata for valuable discussions.
This work was supported by the 21 COE program on ``Center for Diversity and Universality in Physics'' from MEXT of Japan, and by Grants-in-Aid for Scientific Research from the Japan Society for the Promotion of Science (JSPS) and MEXT.


\begin{thebibliography}{99}
\bibitem{MLH}  
$1/T_1T$ was also measured at 12.1 MHz for the MLH sample, and $1/T_1T$ was 8.2 s$^{-1}$K$^{-1}$ at 4.2 K, 10 \% smaller than $1/T_1T$ at 12.3 MHz. Temperature dependence was the same as that at 12.3 MHz.
\end{thebibliography}
\end{document}